\documentclass[manuscript]{aastex}

\slugcomment{} 
\shorttitle{$\eta$ Car -- Mid-Cycle Changes}
\shortauthors{Martin et al.}  

\begin{document}


\title{ Mid-Cycle Changes in Eta Carinae\altaffilmark{1,2,3,4,5} }


\author{John C. Martin}
\affil{Physics and Astronomy Department, University of Illinois, Springfield, IL 62703}
\author{Kris Davidson}
\author{Roberta M. Humphreys}
\and 
\author{Andrea Mehner}
\affil{Astronomy Department, University of Minnesota,
    55455}

\altaffiltext{1}{This research was supported by  
  grants no. GO-10844 and 11291 from the Space Telescope Science Institute.  The HST is
  operated by the Association of Universities for Research in
  Astronomy, Inc., under NASA contract NAS5-26555.} 
\altaffiltext{2}{Some of the data presented in this paper were
  obtained from the Multi-mission Archive at the Space Telescope
  Science Institute (MAST). STScI is operated by the Association of
  Universities for Research in Astronomy, Inc., under NASA contract
  NAS5-26555. Support for MAST for non-HST data is provided by the
  NASA Office of Space Science via grant NAG5-7584 and by other grants
  and contracts.}
\altaffiltext{3}{Based on observations obtained at the Gemini Observatory 
(acquired through the Gemini Science Archive), which is operated by the 
Association of Universities for Research in Astronomy, Inc., under a 
cooperative agreement with the NSF on behalf of the Gemini partnership: 
the National Science Foundation (United States), the Science and 
Technology Facilities Council (United Kingdom), the National Research 
Council (Canada), CONICYT (Chile), the Australian Research Council 
(Australia), MinistB.rio da CiB0ncia e Tecnologia (Brazil) and SECYT 
(Argentina).}
\altaffiltext{4}{Uses data from observations made with ESO Telescopes at the La Silla Paranal Observatory under programme ID 077.D-0618(A)}
\altaffiltext{5}{Uses data from the AAVSO International Database.}


\begin{abstract}
In late 2006, ground-based photometry of $\eta$ Car plus the Homunculus 
showed an unexpected decrease in its integrated apparent brightness, an 
apparent reversal of its long-term brightening. Subsequent HST/WFPC2 
photometry of the central star in the near-UV showed that this was not 
a simple reversal. This multi-wavelength photometry did not support 
increased extinction by dust as the explanation for the decrease in 
brightness.  A spectrum obtained with GMOS on the Gemini-South telescope, 
revealed subtle changes mid-way in $\eta$ Car's 5.5 yr spectroscopic cycle 
when compared with HST/STIS spectra at the same phase in the cycle. 
At mid-cycle the secondary star is 20--30 AU from the primary.  
We suggest that the spectroscopic changes 
are consistent with fluctuations in the density and velocity of the 
primary star's wind, unrelated to the 5.5 yr cycle but possibly related 
to its latitude-dependent morphology.  We also discuss subtle effects that 
must be taken into account when comparing ground-based and HST/STIS spectra. 

\end{abstract}



\section{Introduction}  

Eta Carinae has a peculiar habit of settling into a photometric trend 
and spectroscopic state long enough for observers to grow accustomed 
to it, and then abruptly changing.    
Dramatic transitions occurred around 1840, 1890, and 1945, see 
\citet{2008AJ....135.1249H} and refs.\ therein;
  and there are hints that another 
  may have begun in the late 1990's.
For almost 50 years after 
the 1940-1950 change, the star plus ejecta brightened at an 
average rate of 0.025 visual magnitude per year  
\citep{1956VA......2.1165O,1967Obs....87..287F, 1974A&A....30..271F, 
1998IAPPP..72...53M,1999AJ....118.1777D, 2005ASPC..332..111M}, 
largely due to expansion of the dusty Homunculus nebula which had been 
ejected around 1840.  Essentially a bipolar reflection or scattering nebula, 
the Homunculus dominated $\eta$ Car's total brightness throughout the 
20th century.  After 1997, however, observations with the Hubble Space 
Telescope (HST) revealed that {\it the central star\/} was now 
brightening far more rapidly 
\citep{1999AJ....118.1777D,2004AJ....127.2352M,2006AJ....132.2717M}.  
Despite brief declines during the 1998.0 and 2003.5 spectroscopic 
events, its average rate from 1998 to 2006 was about 0.15 mag/yr.  
This  was a recent development, 
since earlier speckle data \citep{gw86,gw88} show that the star had not 
brightened that quickly in the fifteen years before 1997.   Ground-based 
photometry of the  Homunculus after 1997 showed an increase that 
was less dramatic, but which exceeded any of the fluctuations seen 
between 1955 and 1995 \citep{1999A&A...346L..33S,1999AJ....118.1777D,
2003IBVS.5477....1F,2004MNRAS.352..447W,2004AJ....127.2352M}. Most likely 
the circumstellar extinction began to decrease rapidly in the 
mid-1990's;  perhaps the rate of dust formation near the star had 
decreased, or dust was being destroyed, or both. 
Either case requires some undiagnosed change in the dense 
stellar wind \citep{2006AJ....132.2717M}.

Thus, suspecting that another ``change of state'' was underway, we 
were surprised when ground-based photometry of $\eta$ Car plus the 
Homunculus showed a 0.3-magnitude {\em decrease} in brightness between 
August 2006 and March 2007 \citep{aavsodata,Lajus09}.\footnote{     
   See the CCD photometry by Georgio DiScala (Australia) and by 
   Raymond W.\ Jones (South Africa) in the AAVSO International 
   Database, http://www.aavso.org/ and  the photometric 
   record from La Plata Observatory (Argentina),    
   http://etacar.fcaglp.unlp.edu.ar/.  In our Figure \ref{photom}, 
   note that the decline may have begun in the season that is unfavorable 
   for observing $\eta$ Car.  It may have started as early as August or 
   as late as December of 2006. } 
The change occurred at all BVRI wavelengths and initially resembled the 
behavior seen during a spectroscopic event (Fig.\ \ref{photom}).  This 
was the largest decline in $\eta$ Car's groundbased light curve 
event in at least the past 50 years, see \citet{vanG99}, 
\citet{1999AJ....118.1777D}, and refs.\ therein.   The integrated 
photometry of the Homunculus and central star vary in parallel 
\citep{vanG04}. Therefore, if this change proved to be a 
reversal of the previous several years' behavior, then one 
might expect the central star to have faded by more 
than half a magnitude in 2006--2007, largely invalidating our 
conjectures about the future trend \citep{2006AJ....132.2717M}.    
Thus it was important to obtain fresh HST photometry of the central 
star.  Fortunately this was already a goal of a small existing program
outlined in Section 2 below.

As usual in $\eta$ Car research, the results differ from any scenario 
proposed beforehand.  The photometric decline soon leveled off, but 
our HST photometry strongly suggests that it was {\it not\/} merely 
a temporary reversal of the previous brightening trend (\S {3}). 
Suspecting that the photometric reversal might signal unforeseen spectroscopic 
changes in the stellar wind,  we obtained data with the Gemini Multi-Object
Spectrometer (GMOS) in 2007 which can be compared with earlier
data from the HST's Space Telescope Imaging Spectrograph (STIS) in 1998--2004, 
and with VLT/UVES spectra from 2006  prior to the photometric decline.
The spectroscopy revealed only subtle 
changes halfway through $\eta$ Car's 5.5-year spectroscopic cycle, possibly 
related to the latitude-dependent morphology of the stellar wind.

Any interpretation of $\eta$ Car's recent behavior must take into account 
its 5.54-year spectroscopic cycle; for general information see 
\citet{ASP332}, \citet{2006ApJ...640..474M}, and many references cited 
therein.  Most authors agree that the 1998.0, 2003.5, and 2009.0 
``spectroscopic events'' probably occurred at periastron passages 
of a hypothetical companion star in a highly eccentric orbit.  At 
times more than 7 months from such an event, the two stars are 15 to 30 AU 
apart -- seemingly too distant for the secondary star to significantly 
influence the appearance of the primary wind in the observations discussed 
here. Although \citet{Kashi08a} have suggested that even at apastron the 
ionizing radiation from the secondary may not be negligible. At least 
{\it in most scenarios proposed so far,\/} one expects the 2006--2007 changes 
to signify structural effects in the primary wind more than wind--wind 
interactions.

In the next section we describe the photometric and spectroscopic 
observations including a description of the differences between 
ground-based and HST/STIS spectra and the complications that must be 
accounted for when comparing them.  In \S \ref{three} we discuss the 
observed variations in $\eta$ Car  in 2006--2007, mid way in its
5.5 yr spectroscopic cycle, and compare them with spectra from the 
same phase in the previous cycle (2000--2002). The last section summarizes 
the observations with a brief discussion of their possible origin.

\section{Observations}  

\subsection{HST Photometry}  

Normal ground-based images of $\eta$ Car do not consistently separate 
the star from the inner parts of the Homunculus nebula, which has 
structure on all scales from 0.1{\arcsec} to 10{\arcsec}.  Until 
recently the Homunculus appeared brighter than the star, though the 
ejecta-to-star brightness ratio has evolved substantially in the past 
20 years \citep{2006AJ....132.2717M}.  Precise ground-based photometry 
generally includes the entire Homunculus plus the star.  HST images,  
however, have good enough spatial resolution to show the brightness of 
the star itself with only minor contamination by the ejecta located at 
$r \gtrsim 0.15{\arcsec}$.  In recent years we have obtained a number 
of images of this type, to monitor the central star's brightness and 
detect changes in the surrounding ejecta.

We obtained Hubble Space Telescope ACS/HRC images of $\eta$ Car in the 
F250W and F330W filters on 2006 August 4 (MJD 53951.1), near peak 
brightness, and 2007 January 20 (MJD 54120.4), during the unexpected 
decline.  The central wavelengths of these broad-band filters are near 
250 and 330 nm respectively.  All of the images were reduced, measured, 
and calibrated  as described in \citet{2006AJ....132.2717M}.   For 
brightness measurements we used a weighted virtual aperture centered 
on the star, with diameter  0.3{\arcsec} (about 10.7 ACS/HRC 
pixels).  The aperture weighting function was $\, 1 - r^2/R^2 \,$ 
with $\,R$ = 0.15{\arcsec}.\footnote{  
   In \citet{2006AJ....132.2717M} we wrongly quoted $\,R$ = 0.3{\arcsec}, 
   which was really the diameter $2R$.   \citet{2004AJ....127.2352M} 
   stated this detail correctly. }  
See also \citet{Smith04} for aperture size effects in ACS images of $\eta$ Car. 

Shortly after our January 2007 observations the ACS electronics 
failed.  We then used the older HST/WFPC2 camera on 2007 August 23 
(MJD 54335.1) 
and 2008 February 14 (MJD 54510.8), with filters F255W and F336W which 
are comparable to the ACS/HRC filters.  These images were reduced 
using the standard STScI data reduction pipeline.  A correction factor 
of 0.9915 was applied to the raw counts to compensate for geometric pixel 
distortion \citep{WFPC2_Handbook}, and the PHOTFLAM keyword induced a 
pipeline conversion from raw counts to flux values.    

We measured the brightness of the central star in the WFPC2 data, using 
the same 0.3{\arcsec} weighted virtual aperture ($\sim$ 6.6 PC2 pixels) 
as for the ACS/HRC data.  We estimated the aperture correction based on 
two 12th-magnitude stars in the images, well separated from the Homunculus,
by comparing their fluxes in the virtual 
aperture to those in a standard 
1{\arcsec} measurement aperture corrected to infinite size 
\citep{1995PASP..107..156H}.  The resulting geometrical transmission 
factors are listed in Table \ref{caltab}. 

To assess the effect of a known ``red leak'' in the WFPC2 F336W filter 
transmission curve, we convolved an  HST/STIS spectrum of $\eta$ Car 
with the response function for the Planetary Camera with this filter.  
The flux at $\lambda > 450$ nm contributed only 0.94\% of the counts, 
negligible for our purposes.  

The STMAG photometric system is calibrated for direct comparison of fluxes 
measured with similar filters in different instruments 
\citep{2005PASP..117.1049S}.  We tested the calibration by convolving the 
instrument and filter response functions with HST/STIS spectra of $\eta$
Car and comparing the results for the ACS/HRC F250W and F330W filters to 
those for the WFPC2 F255W and F336W filters.  The computed values 
agreed to four significant digits, giving us confidence that comparisons 
between our ACS/HRC and WFPC2 results are valid. 
Our experience with the 
large existing HST data set for $\eta$ Car, 1991--2009, has never 
contradicted this assessment to a serious degree, despite the spectral  
forest of emission lines and strong UV absorption.

The ACS/HRC and WFPC2 photometry are listed in Table \ref{phottab}, 
and plotted in Figure \ref{photom} together with our previous 
measurements  \citep{2006AJ....132.2717M}.

\subsection{Spectroscopy \label{specsec}}   

\begin{figure}   
   \includegraphics[angle=270,scale=0.7]{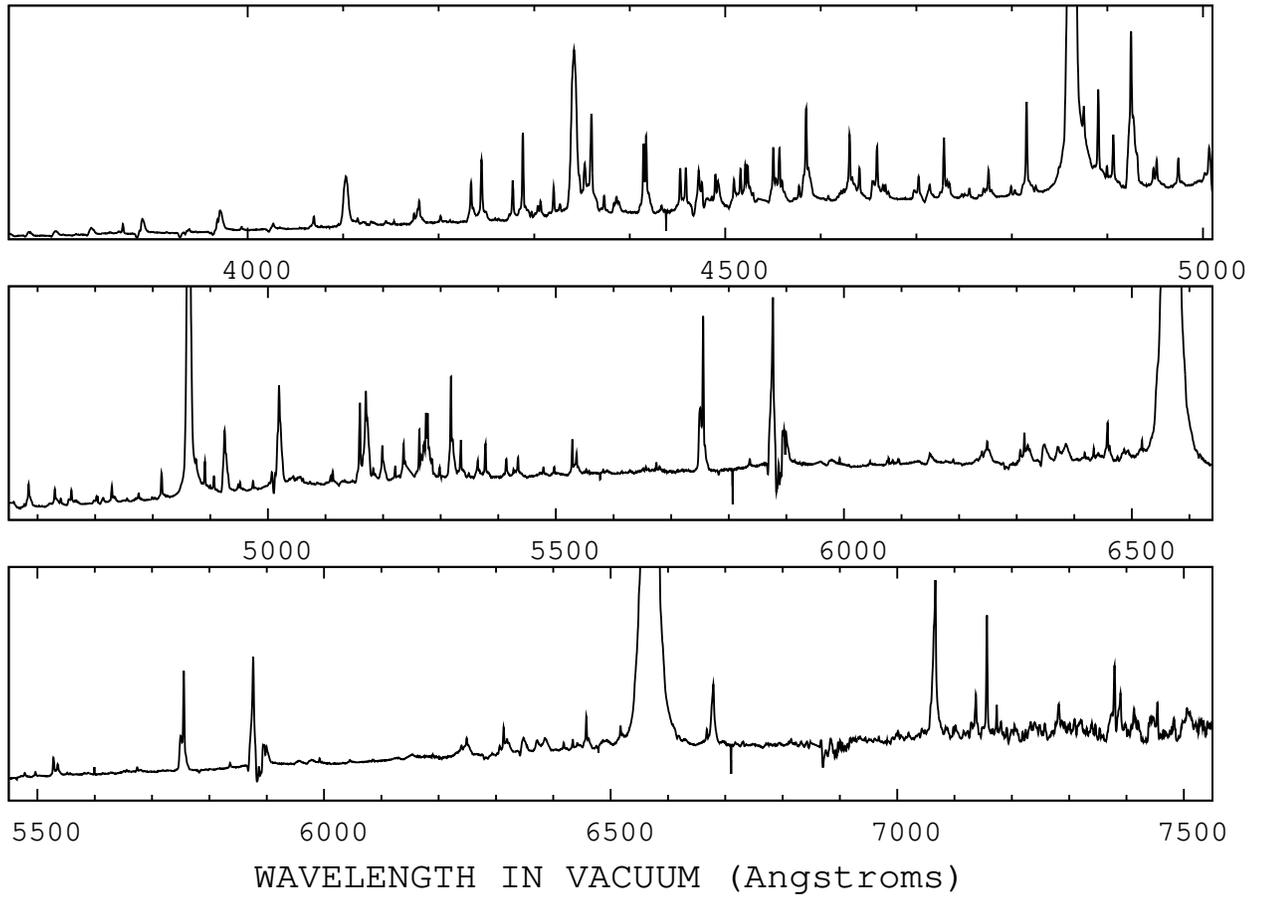}
   \caption{A tracing of the entire GMOS spectrum of the star 
      in July 2007.}
   \label{fullspec}
   \end{figure}

We obtained ground-based slit spectroscopy of $\eta$ Car  with the Gemini 
Multi-Object Spectrograph (GMOS) on the Gemini South telescope in late June 
and early July 2007, just after the decrease in brightness.  
The parameters of these observations and subsequent 
GMOS observations in February and July 2008
are listed in Table \ref{spectab}.  The parameters of the 
spectrograph are given in Table \ref{instab}.  We prepared 
2-D spectrograms with the standard GMOS data reduction pipeline in the 
Gemini IRAF package and
extracted 1-D spectra via a routine 
developed earlier for use with HST/STIS \citep{2006ApJ...640..474M}
(Figure \ref{fullspec}). 
At each wavelength our software integrates the counts 
along a line perpendicular to the dispersion, weighted by a mesa-shaped 
function centered on the local spectral trace.  For these data we used 
a mesa function with base-width = 11 pixels and top-width = 7 pixels,
about 0.8{\arcsec} and 0.5{\arcsec} respectively.  The seeing was 
roughly 1.0{\arcsec}, so each GMOS spectrum discussed in {\S}3 
below represents a region about 1{\arcsec} across.  

HST/STIS spectra of $\eta$ Car had been obtained in October 2001 and 
January 2002, at approximately the same phase in its 5.54-year spectroscopic 
cycle \citep{Davidson04}.  We reduced them using a modified version 
of the Goddard CALSTIS reduction pipeline 
described in  \citet{2006ApJ...640..474M}.  
The spectra are listed in Table \ref{spectab} and the instrument
parameters are given in Table \ref{instab}.  The January 2002 data 
most closely match the spectroscopic phase of July 2007.\footnote{   
      STIS and GMOS spectra used in this paper can be downloaded 
      at http://etacar.umn.edu/.  } 

  Of course the spatial resolution 
of Gemini, limited by atmospheric seeing, is greatly inferior to 
HST/STIS and the inner ejecta are unavoidably
mixed with the spectrum of the star.  
In particular our GMOS spectra include the ``Weigelt knots'' 
0.3{\arcsec} northwest of the star.  Fortunately, by 2007 the star 
had become substantially brighter than the nearby ejecta -- unlike the 
case 10 or 20 years earlier  \citep{2006ApJ...640..474M}.    
The 1-D STIS spectra used in this paper were prepared 
with an extraction width of 0.20{\arcsec}, which excludes most of the 
narrow emission from the inner ejecta;   see important comments in 
subsection \ref{two.three}.      
  
Moreover, the slow-moving inner ejecta produce narrow emission lines 
which are distinguishable from the broad stellar wind lines; 
typical widths are of the order of 20 and 400 km s$^{-1}$ respectively.  
In the wavelength range of interest, the spectral resolution of GMOS 
is roughly 75 km s$^{-1}$ while STIS provided about 40 km s$^{-1}$.   
The narrow lines are therefore  more blurred in the Gemini data
while the broad stellar wind features and their 
P-Cygni absorption components are well resolved by both instruments. 
We have smoothed the STIS data plotted in Section \ref{three} to 
match the resolution of GMOS.    

H$\alpha$ emission is so bright in $\eta$ Car that it saturates the 
detector pixels in even the shortest GMOS exposures centered on the star.
We attempted to use a narrower slit, offset from the star by 0.6{\arcsec} 
to sample the wings of the star's p.s.f.   This technique produced 
an unsaturated H$\alpha$ profile, but it is probably too contaminated 
by surrounding ejecta to be trusted;   see Subsection \ref{two.three}.

Unfortunately $\eta$ Car was not observed with Gemini/GMOS in 2005--2006
before the photometric fading.   
The ESO Science Archive does have spectra obtained with the VLT/UVES 
instrument in 2006\footnote{ 
      The 2006 UVES observations were obtained at ESO's  Paranal 
      Observatory under program ID 077.D-0618(A) by Weis et al. 
      VLT/UVES observations of $\eta$ Car from 2002 to 2005 are 
      available at http://etacar.umn.edu/. The UVES spectral sequence is also described
      by \citet{Nielsen09}, but they did not note the potential problems with the later 
      UVES spectra.}.   
We reduced the 2006 spectra with the standard UVES pipeline available from ESO.
The spectra were extracted using a mesa function 3 by 2 pixels wide, 
about 0.75$\arcsec$ by 0.5$\arcsec$. 
The seeing was 0.8 to 0.9$\arcsec$ so this extraction corresponds to 
about 1$\arcsec$ on the sky.
None of the 2006 spectra are concurrent with the decrease. The closest in time is from June
2006 but internal clues suggest that the slit was not centered on the star. 
The spectra  do not closely resemble earlier UVES observations (2002 - 2004)  
which were definitely centered on the star; the stellar-wind features 
are weaker and the narrow emission lines are much weaker than 
what one expects to see in a ground-based spectrum. The published position and 
acquisition image for the June 2006 spectrum also suggest that the slit might be offset from 
the star.
The June 2006 UVES spectrum thus appears unsuitable for our purposes here. 
The  narrow emission lines are also weaker than expected in the two
spectra from 2005 and the April 2006 spectrum and weaker 
than in the 2007 GMOS spectra. This may also be due to 
pointing differences with respect to the earlier UVES spectra and the GMOS spectra. 
Although the April 2006 spectrum is referenced in the later discussion, 
it is used here with reservations for this reason.

\subsection{An unsolved puzzle:  The effect of spatial resolution 
    on spectroscopy of $\eta$~Car\label{two.three}}  

Provided the extraction size is less than 0.3{\arcsec}, an HST 
spectrum of $\eta$ Car shows the stellar wind with only slight  
contamination.  Ground-based spectroscopy, on the other hand, with seeing 
of the order of 1{\arcsec},  also includes the spectrum reflected 
by dust at $r \sim $ 0.15{\arcsec} to 1{\arcsec}, plus narrow emission 
lines created in that region.   Since reflection usually has little effect 
on the strengths of spectral features relative to the continuum, 
one   might expect a ground-based spectrum of $\eta$ Car (i.e., 
central object plus reflected spectrum plus narrow-line emission) to 
closely resemble an HST spectrum, supplemented by the narrow lines.  
In fact, however, the earliest HST spectra showed that {\it this is 
not the case} \citep{fos95}.  Ground-based data show paradoxically  
weaker Balmer emission lines.

\begin{figure}  
  \includegraphics[angle=0,scale=0.25]{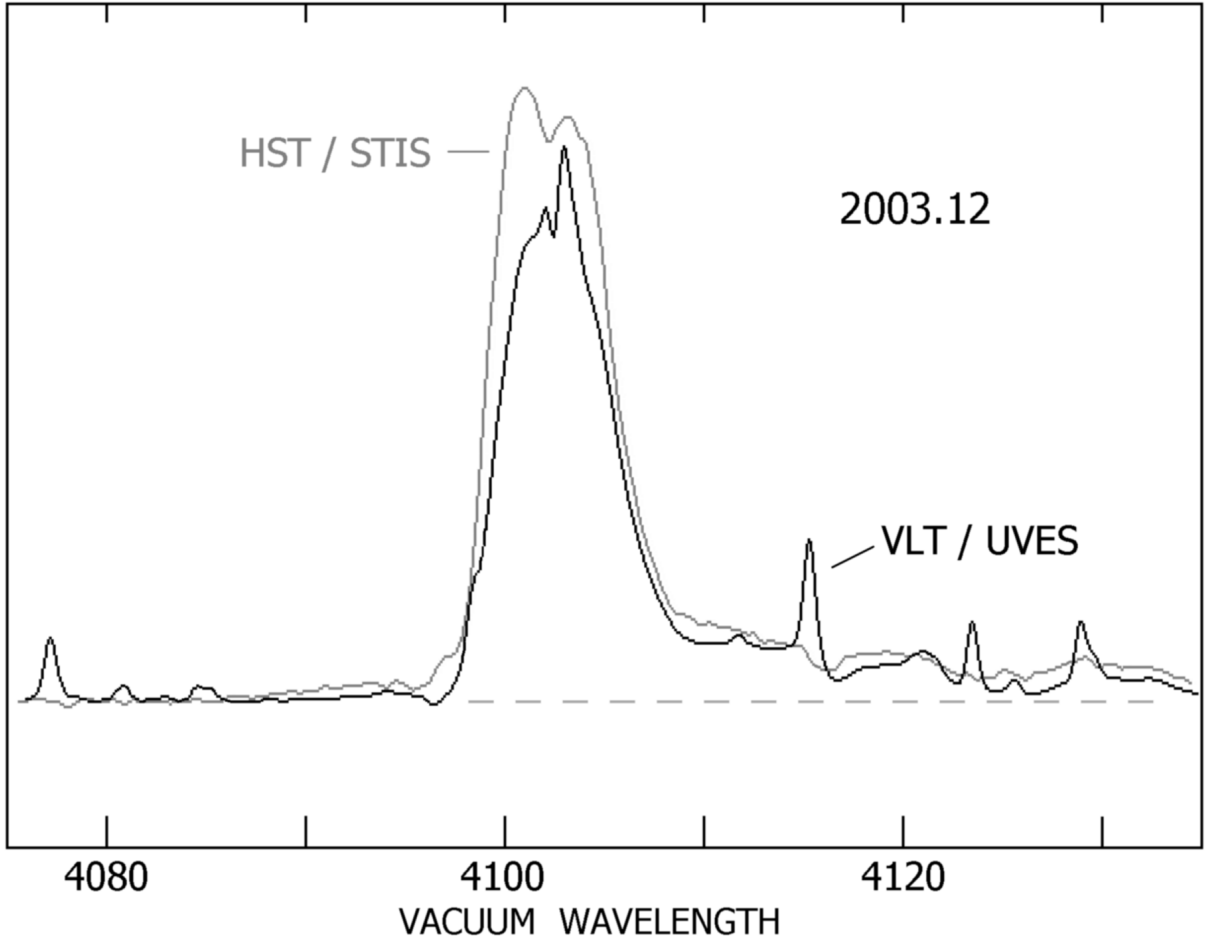}   
  \caption{The H$\delta$ profile in $\eta$ Car, observed with both 
    HST/STIS and VLT/UVES at practically the same time in February 2003. 
    The STIS tracing samples a diameter less than 0.2{\arcsec},  
    while the UVES spectrum represents a seeing-limited region 
    about 1{\arcsec} across.  Both data sets have been normalized 
    so their continuum fluxes match near $\lambda \approx 4080$ {\AA} 
    and 4160 {\AA}, and the bottom edge of the figure corresponds to 
    zero flux.   The spectral resolution of the UVES tracing has been 
    degraded to match that of STIS, roughly 40 km s$^{-1}$. }        
  \label{stisuves}        
  \end{figure}

Fig. \ref{stisuves} shows how conspicuous this effect is.  Here 
we use VLT/UVES to exemplify a modern ground-based instrument, 
because $\eta$ Car was observed with STIS and UVES on 2003 February 
12 and 14 respectively, only 1.2 days apart.  We show the H$\delta$ 
feature because it plays a leading role in \S {3.2}.  
H$\delta$ emission in the UVES data 
has  equivalent width 26 {\AA} compared to 33 {\AA} in 
the STIS spectrum 
(measured the same way in both data sets, with 
no correction for narrow lines).  Old HST/FOS results and various 
ground-based data suggest that this difference is real, not merely an 
instrumental peculiarity of either STIS or UVES \citep{fos95}.

Why does the difference occur?  Wavelength-dependent Balmer absorption 
within the reflecting dust-gas mixture seems implausible, because each 
stellar-wind emission line is much broader than the velocity dispersion 
in the relevant material.  A more appealing explanation is that the 
spectrum ``seen by'' the dust intrinsically differs from what we see directly 
--  in other words, the stellar wind spectrum depends on the direction 
from which it is viewed. \citet{2003ApJ...586..432S} explored this idea 
via observations farther out in the Homunculus Nebula, and concluded 
that $\eta$ Car's wind is strongly latitude-dependent.   Note that 
the dusty ``Weigelt knots,'' which contribute strongly to any 
ground-based spectrum of the central object,  are thought to be near 
the equatorial plane while our direct viewpoint is around 45$^{\circ}$ 
latitude e.g. Davidson et al (2001).    

This explanation is not entirely satisfying. 
It implies 
that the equivalent widths of Balmer lines are substantially 
smaller at equatorial latitudes, whereas \citet{2003ApJ...586..432S} 
and \citet{ha92} suggested the same for {\it polar\/} directions.   
Apparently HST's high-spatial-resolution direct view of the stellar 
wind gives larger equivalent widths than we find in the reflected 
spectrum anywhere in the ejecta.   (This statement probably applies
to broad emission features in general, not just Balmer lines.) 
Furthermore, the UVES vs.\ STIS comparison mentioned above seems to 
require a surprisingly large latitude variation in the H$\delta$ 
equivalent width, almost a factor of two. Thus we fear that some 
other, more subtle process may be altering the reflected spectra.  
This problem merits further study.    

Independent of what causes the discrepancies, for this paper the 
practical implication is that {\it one must be very careful when 
comparing ground-based spectra of $\eta$ Car to the HST/STIS data.\/} 
In particular we cannot directly compare the equivalent widths.   
Moreover, the offset H$\alpha$ observations mentioned in {\S}2.2 
are not useful here because they are most likely affected by reflection 
and emission in their spatial locales. In \S {4} below we compare 
recent Gemini/GMOS observations to earlier HST/STIS results, because 
neither data set spans two spectroscopic cycles;  nor is any other  
existing data set suitable for this problem.  In those comparisons, 
we exercise the necessary caution and caveats.

\section{Mid-Cycle Activity\label{three}}   

In recent years relatively little attention has been given to 
$\eta$ Car's behavior between spectroscopic events.  
At such times the companion star must be located 15--35 AU 
from the primary.  Mid-cycle 
changes at UV-to-red wavelengths are therefore significant for at 
least one of two reasons:  (1) Proposed explanations for the 5.54-year 
cycle generally do not predict irregular or short-term effects more 
than a few months from an event.  If they do occur and are related to 
the cycle, then some new factor must be taken into account. (2)  If, on 
the other hand, such effects are not related to the cycle, then they 
may represent phenomena in the primary wind.  Observations in 1999--2002 and 
2005--2007 therefore merit attention.  Here we are concentrating on 
2006--2007, supplemented by a few data from 2000--2002 for comparison 
purposes.

  To avoid misunderstandings, note that continuous {\it systematic\/} 
changes occur in mid-cycle, not directly related to the primary star.    
\citet{mehner2010} have shown that some quasi-nebular spectral lines 
emitted close to our line of sight to the star gradually brighten 
to a maximum near mid-cycle and then fade.   \citet{ad2008} earlier found 
a similar result for one line of this type in an unresolved set of emission 
regions.   The spectral features in question ([\ion{Ne}{3}], [\ion{Ar}{3}], 
[\ion{Fe}{3}], etc.) probably indicate photoionization by the hot secondary 
star, and their behavior in the Mehner et al.\ analysis may represent the 
cyclic variation of column density between us and the orbiting secondary 
star.  The most popular class of orbit-and-wind models can probably be 
adjusted to match this effect.   Fluctuations described below, however, 
differ in several respects.  They directly concern the spectrum 
of the primary stellar wind rather than nearby ejecta;  they don't have 
a smooth obvious correlation with phase in the 5.5-year cycle;  and 
each of them occurred in only one of two observed cycles.

\subsection{Photometry in 2006--2007}   

The HST images confirm that the central star became fainter during 
2006--2007 (Fig.\ \ref{photom}),  but not as much as the ground-based 
photometry had led us to expect.  From 1997 to 2006 the star's apparent 
magnitude had brightened three times faster than the surrounding Homunculus,
but {\it its relative fading in 2006 was only about the same as for the 
Homunculus.\/}  Whatever the explanation is, this detail indicates 
that the change in 2006--2007 was not a simple reversal of the 
1998--2006 trend.  Moreover, the decline stopped
and the upward brightening trend resumed after the
2009 spectroscopic event, leaving more or 
less intact our earlier speculations about secular changes 
\citep{2006AJ....132.2717M,2005AJ....129..900D}.

The F330W filter showed a much steeper decline than F250W.  This is 
interesting because the former samples Balmer continuum emission plus 
other continuum, whereas F250W includes many strong 
\ion{Fe}{2} absorption features.  
During a spectroscopic event  such as 1998.0 or 2003.5, these absorption 
lines strengthen so much that the star becomes faint in the 230--290 nm 
wavelength range \citep{cassatella79,altamore86,viotti89,stis99}.    
This is why Fig.\  \ref{photom} shows a substantial temporary decrease 
in F250W during the 2003.5 event,  with scarcely any corresponding effect 
in F330W.  Evidently the 2006-2007 fading was quite different. 
The same  figure suggests two possibilities:     
\ (1) The {\it increased\/} F250W/F330W flux ratio in 2007 does not
favor an explanation based on increased extinction by circumstellar dust.  
\ (2) Perhaps the Balmer continuum emission intrinsically weakened in 2006.  

The ground-based B, V, R, and I magnitudes all faded by about 
$\Delta m \approx +0.3$ in 2006--2007 \citep{aavsodata}.  
While $\eta$ Car's B and R values are strongly influenced by the 
very bright H$\beta$ and H$\alpha$ emission lines, V almost entirely 
measures continuum brightness.  Unfortunately the central star could 
not be measured with HST/WFPC2 at these wavelengths, because 
every WFPC2 filter redward of 400 nm either saturates the central 
pixels in the shortest allowed exposure time, or else is affected 
by some particular emission line.

\subsection{Spectroscopy: 2007 vs.\ 2000--2002}  

Motivated by the visual decline, its duration, and the peculiar 
UV photometry from HST/WFPC2,  we obtained the
Gemini/GMOS data described in {\S}\ref{specsec} to learn whether the spectrum had 
changed during the brightness decrease.  Unfortunately  no 
truly suitable earlier ground-based spectra are available for
comparison -- particularly none from 2005 or 2006.  
Instead, therefore, we used STIS observations 
made in October 2001 and January 2002, at approximately the same 
phase of $\eta$ Car's 5.54-year spectroscopic cycle as the GMOS data.  
There is no {\it a priori\/} reason to think that the 2006 photometric 
change was related to the cycle, but on the other hand it would be 
imprudent to use observations at much different phases as comparisons. 
If, as usually assumed, the cycle is modulated by a 
hot companion star in an eccentric orbit, then Fig.\ \ref{orbitfig} 
shows the approximate orbital location of the secondary star when 
our spectra were obtained.\footnote{  
    Most of the orbital parameters are extremely uncertain, but 
    Fig.\ \ref{orbitfig} is qualitatively valid for nearly all 
    models that have been proposed.  See \citet{ASP332,kiorbit2001,
    aoorbit2008} and refs.\ therein.} 
In this part of the  orbit the two stars are more than 20 AU apart and 
their motion is slow.

\begin{figure} 
  \includegraphics[angle=0,scale=0.70]{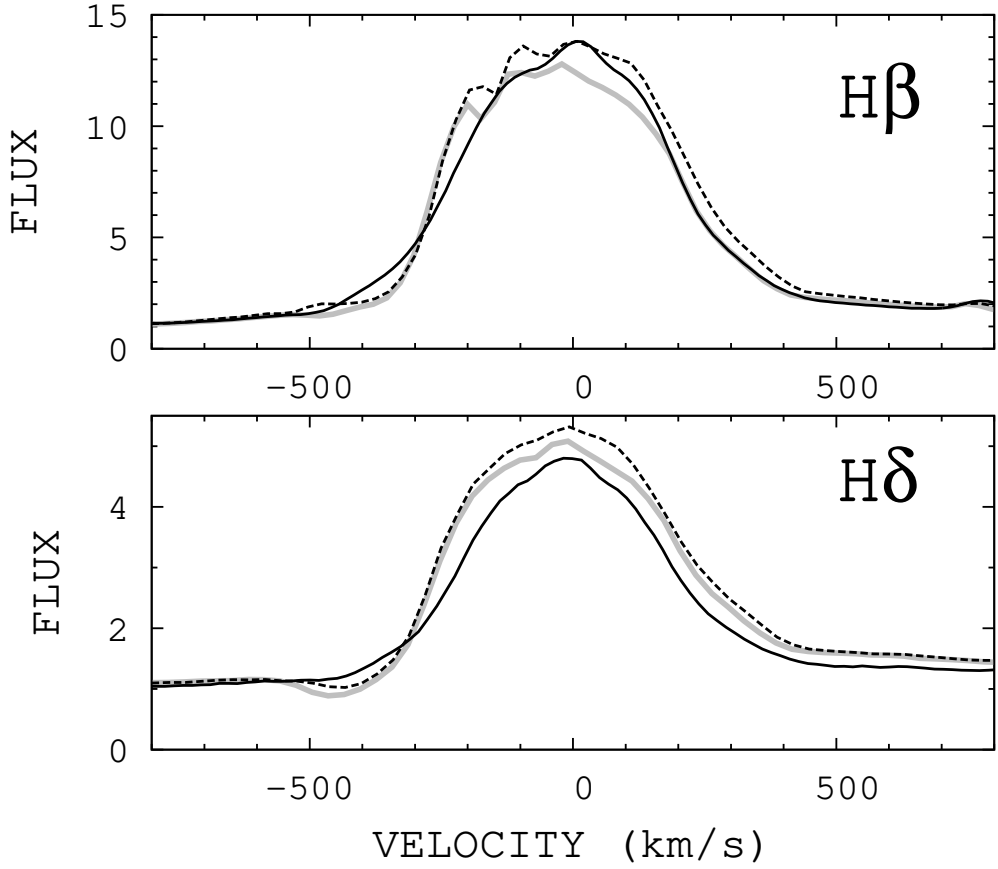}
  \caption{Line profiles of two bright Balmer lines.  The June 2007 
     GMOS spectrum is the solid black line.  The solid gray and dashed black 
     lines are the STIS spectra from October 2001 and January 2002 
     respectively, smoothed to match the GMOS resolution of about 
     75 km s$^{-1}$.}
  \label{balmer}
  \end{figure}

The most conspicuous spectral features in 2007 (Fig.\ \ref{fullspec}) 
closely resembled the 2001--2002 data. Consider for instance the 
brightest Balmer lines.  Since H$\alpha$ is overexposed in the GMOS 
data and H$\gamma$ is confused with other features in this object,  
in Fig.\ \ref{balmer} we show H$\beta$ and H$\delta$.
The measured equivalent widths for H$\delta$ turn out to be 
like those mentioned in {\S}2.3:  
26 {\AA} in the 2007 GMOS spectrum and 33 {\AA} 
in the 2001--2002 STIS data.  
Unlike the earlier examples, however, 
these values have been corrected for weak superimposed narrow lines.  
The internal uncertainties -- chiefly systematic rather 
than statistical -- are probably less than $\pm$ 2 {\AA}.\footnote{  
   Since the continuum of $\eta$ Car is hard to define and 
   weaker additional emission lines contribute to the 
   fluxes, these results depend on the measurement protocol. We measured 
   ``continuum'' fluxes near 4080 and 4160 {\AA} and we integrated the 
   net H$\delta$ flux from 4082 to 4116 {\AA} (vac).  The correction 
   for unrelated emission on the 
   long-wavelength wing is small, about $-0.6$ {\AA} in the GMOS data 
   and less for STIS.  In the last STIS data in early 2004 the equivalent 
   width for H$\delta$ was only about 29 {\AA}, but lingering effects of 
   the 2003.5 spectroscopic event may have reduced it below 
   ``normal'' \citep{2005AJ....129..900D}.  Regarding earlier epochs, 
   the first HST spectra gave an E.W. of 35 {\AA} for H$\delta$ in 1991 
   \citep{fos95}, or probably about 33 {\AA} if we could apply the 
   corrections mentioned above. } 
If we make allowances for the effect described in {\S}2.3, evidently 
H$\delta$ was essentially as strong in 2007 (relative to the continuum) 
as it had been 5.5 years earlier \footnote{Following the same protocol, the equivalent 
widths for H$\delta$ in the February and
July 2008 GMOS spectra are 25 {\AA} each and 26 {\AA} in the April 2006 UVES spectrum.}.
The H$\beta$ equivalent width was about 163, 175, and 156 {\AA}, 
respectively, in the 2001 and 2002 STIS data and the 2007 GMOS data; 
this line may be less accurate than H$\delta$ because its very 
large line/continuum flux ratio is disadvantageous for the measurements.

\begin{figure} 
  \includegraphics[angle=0,scale=0.70]{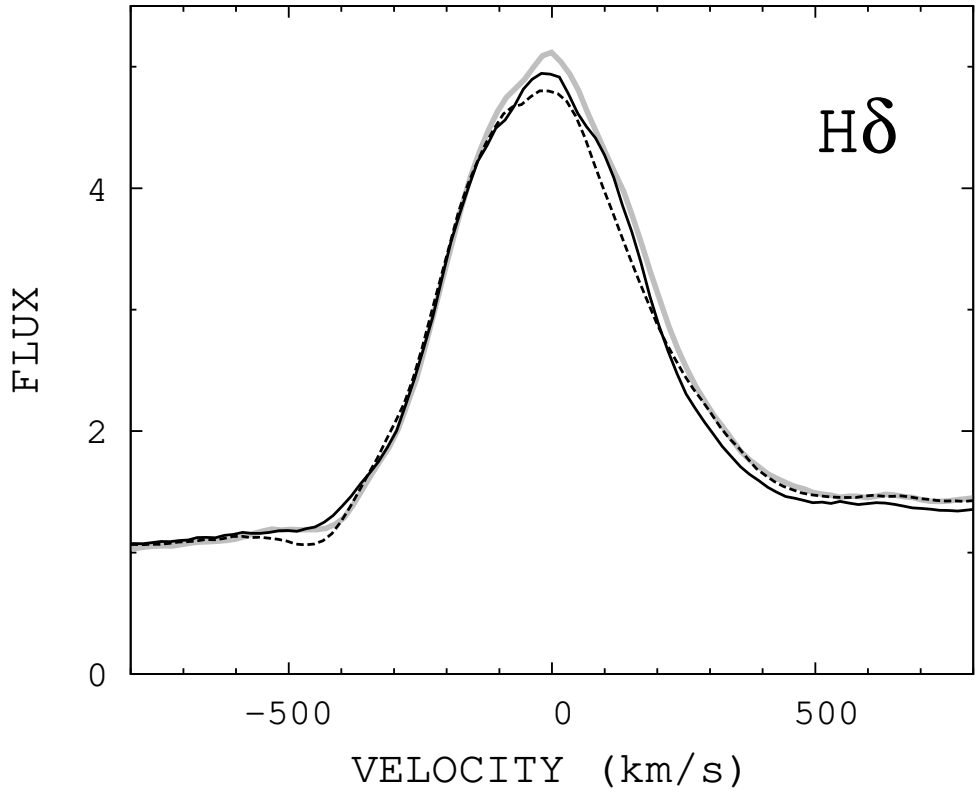}
  \caption{The H-delta line profiles for GMOS spectra observed July 2007
  (solid black line), February 2008 (solid gray line), and July 2008
  (dashed black line).  Note that P-Cygni absorption is present in only 
  the July 2008 spectrum.}
  \label{gmoshdel}
  \end{figure}

But the H$\delta$ profile does show one potentially significant 
difference: Weak P-Cygni absorption was present near $-450$ km s$^{-1}$ 
in 2001--2002 and in the April 2006 UVES spectrum, but not in 2007.  
This is not merely a result of different spectral resolutions, 
because the STIS data in Fig.\ \ref{balmer} have been smoothed to 
match the GMOS resolution.  The difference in spatial resolution is 
probably not responsible either, since the STIS vs.\ UVES comparison 
in Fig.\ \ref{stisuves} shows at most a very weak P-Cygni feature 
in the ground-based data rather than STIS.  (For reasons why that  
might occur, see \cite{2003ApJ...586..432S} along with {\S}2.3 above.)   
The GMOS spectrum and H$\delta$ profile from February 2008 is essentially identical to the 
2007 spectrum, while  its July 2008 spectrum shows the reappearance of weak 
P Cygni absorption in H$\delta$ prior to the onset of the 2009 event 
(Figure \ref{gmoshdel}).  Although the spectroscopic analysis by \citet{Hill01}
pertained to the HST/STIS spectrum only one year after the 1998.0 event, they
showed that the hydrogen emission originates predominantly in the outer wind of the primary star.  Altogether, then, the P-Cygni difference is a real effect, 
meaning that the column density of excited hydrogen atoms along our 
line of sight was less after the photometric decline.

A parallel, but more dramatic, difference occurred in some of the helium 
features.  Helium lines whose lower levels are 1s2s $^1$S, 1s2s $^3$S, or 
1s2p $^3$P$^\mathrm{o}$ show prominent P-Cygni absorption in $\eta$ Car's
mid-cycle spectrum;   good examples are the triplet features at 
$\lambda\lambda$4026,4472,4713.  Qualitatively, one expects such 
absorption to occur in a dense photoionized He$^+$ zone where 
recombination strongly populates 
the metastable 1s2s levels \citep{ofbook05}.  Such a zone is thought to 
exist in those parts of $\eta$ Car's primary wind that are closest to the 
hot secondary star before encountering the wind-wind shocks;  see 
Section 6 of \citet{2008AJ....135.1249H} and refs.\ therein.     
If Fig.\ \ref{orbitfig} is more or less valid, that zone was situated 
between us and the primary wind in 2001-2002 and 2007;  so P-Cygni 
absorption is unsurprising.  For our purposes here, the interesting 
point is that {\it the helium P-Cygni velocities differed between those 
two cycles.}

\begin{figure}  
\figurenum{7a}
  \includegraphics[angle=0,scale=0.70]{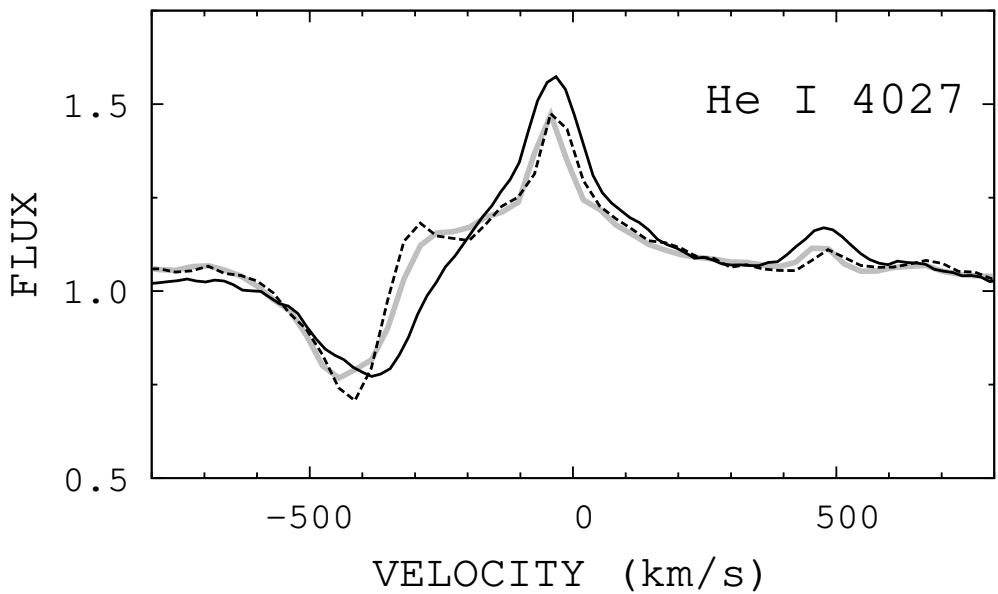}
  \caption{A comparison of the velocity structure in the He I $\lambda$4027 
    triplet line.  As in Figure \ref{balmer}, the June 2007 
    GMOS spectrum is the solid black line.  The solid gray and dashed black
    lines are the STIS spectra from October 2001 and January 2002 respectively
    smoothed to the same resolution as the GMOS spectrum.}
  \label{hebalmer}
\end{figure}

\begin{figure}  
\figurenum{7b}
  \includegraphics[angle=0,scale=0.70]{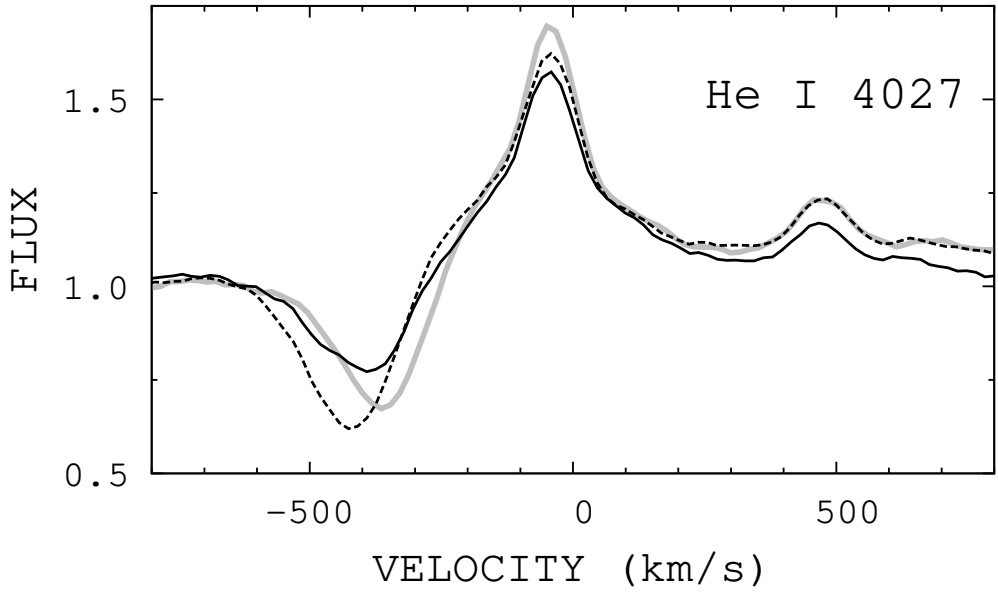}
  \caption{A comparison of the velocity structure in the He I $\lambda$4027 
    triplet line in the GMOS spectra from July 2007 (solid black line), 
    February 2008 (solid gray line), and July 2008 (dashed black line).
    Note the blueward shift of the P-Cygni absorption in the July 2008 
    spectrum.
  }
  \label{heallgmos}
\end{figure}

\begin{figure}  
\figurenum{7c}
  \includegraphics[angle=0,scale=0.70]{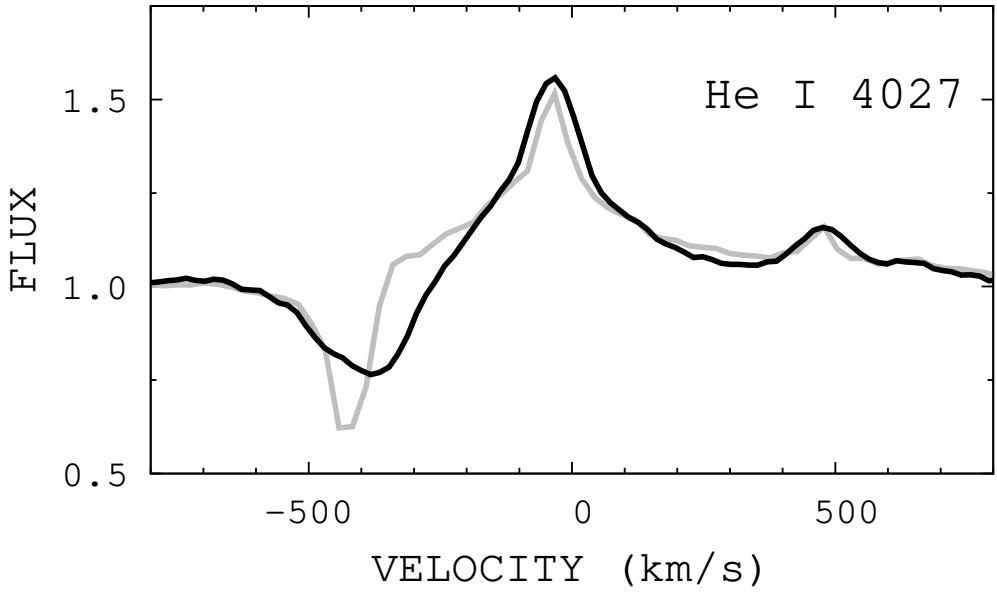}
  \caption{A comparison of the velocity structure in the He I $\lambda$4027 
    triplet line in the GMOS spectra from July 2007 (solid black line) and
    the UVES spectrum from April 2006 (solid gray line).  The UVES data has 
    been binned by a factor of five to smooth out noise but it is still plotted
    with five times greater resolution than the GMOS spectra.  Note that
    the P-Cygni absorption in the 2006 UVES spectra is asymmetric and has 
    two velocity components.
  }
  \label{hegmosuves}
\end{figure}

Fig.\ \ref{hebalmer} shows \ion{He}{1} $\lambda$4027.  In 2001--2002 its 
P-Cygni feature was centered at $V \sim -450$ km s$^{-1}$ like that 
of H$\delta$, but in 2007 the absorption had broadened and shifted 
 by about 100 km s$^{-1}$ to smaller (i.e., less negative) average 
velocities.  This most 
likely indicates a quantitative change in the primary wind structure.
Similar effects occurred in the other helium lines. The P~Cygni absorption 
is also stronger in this line than the corresponding emission. We also 
note that the He I profile has continued to shift to even lower velocities in  
the February 2008 GMOS spectrum, but  it has shifted bluewards, back to 
$\sim -450$ km s$^{-1}$, in the July 2008 spectrum (Fig \ref{heallgmos}). 
Interestingly, the \ion{He}{1} $\lambda$4027 P Cygni profile in the April 
2006 UVES high resolution spectrum appears to have two absorption 
components centered at  velocities of $\sim -430$ and $\sim -400$ km s$^{-1}$ 
(Fig \ref{hegmosuves}).  It is uncertain however, if the UVES slit is 
centered on the star or intercepting the same  scattered and reflected 
light from the surrounding ejecta as the GMOS spectra. 
We plan to model the behavior of the He I lines during $\eta$ Car's 
spectroscopic cycle, but this is beyond the scope of this paper.

By now it is clear that $\eta$ Car is not a simple 5.54-year clock-like  
mechanism.  HST/STIS data show a partially different set of changes 
in the previous mid-cycle interval, 2000--2001 \citep{2006PASP..118..697M}.  
For example, P Cygni absorption in H$\alpha$, usually absent in a mid-cycle 
spectrum of the star \citep{2003ApJ...586..432S}, showed a marked increase 
in absorption between March 2000, when there was no detectable P Cygni 
absorption, and October 2000. Its appearance was brief since it was
gone by January 2001.   Unfortunately the October 2000 STIS data only 
covered H$\alpha$ so we could not compare the profiles of any of the other 
Balmer features or He I triplet lines.  At the same time, there was a 
significant increase in the strength of the narrow -140 km s$^{-1}$ 
absorption feature in H$\alpha$.  The lower spectroscopic resolution of 
GMOS makes it difficult to determine if the narrow $-140$ km s$^{-1}$ feature 
was present in 2007.

 The variable  emission feature near 6307 {\AA} reaches its maximum strength  
 mid-cycle, in 2001--2002 and again in 2005 \citep{2006PASP..118..697M}. 
 In the  VLT/UVES spectra from 2005 it had an equivalent width more than twice 
 that previously observed. 
 It is also present in the April 2006 UVES spectrum and June 2007 GMOS spectrum 
 at 0.2 -- 0.3{\AA} equivalent width\footnote{
 As outlined in  \citep{2006PASP..118..697M} this line blends with an adjacent
 emission feature so it is difficult to accurately measure its strength 
 without a more in depth analysis including spectra taken during
 a spectroscopic event.}, equal to its typical strength reported 1999--2002
 between spectroscopic events.  It is
 slightly stronger in the April 2006 spectrum relative to the June 2007 
 spectrum.  We note that an increase in the UV flux from
 the central star occurs in 2001--2002 and 2005--2006 and corresponds with 
 the maximum strength of the 6307{\AA} line.  

There was at least one episode of mid-cycle activity in each of the 
last two spectroscopic cycles.   Each episode was 
different and each occurred at a different phase relative to the 
spectroscopic events.  In the next section we summarize the mid-cycle 
variability and discuss probable variations in the primary star's  wind  
that would account for these changes. 


\section{Discussion}\label{discussion}   

The mid-cycle  variability in 2006-2007
 described in {\S}3 occurred when the secondary star was at its greatest distance from the 
primary and most likely reveals basic fluctuations 
of $\eta$ Car's primary stellar wind, independent of the companion 
star and the 5.5-year cycle.  This is the {\it simplest\/} view for reasons 
outlined below, and it makes sense in relation to several known or 
probable facts.    For the first time in 160 years, the central star now 
strongly and directly affects ground-based photometry.   The 
Homunculus nebula dominated the total apparent brightness throughout 
the 20th century,  see refs.\ in \citet{2006AJ....132.2717M};  but since 
1998  the star has been brightening rapidly and now accounts for 
 half or more of the integrated light.  This development may explain why ground-based 
magnitudes (i.e., Homunculus plus star) have recently been fluctuating 
with greater amplitude than in the 1953--1995 record. See for example, the recent light
curve at http://etacar.fcaglp.unlp.edu.ar . 
Apparently our direct view of the star shows stronger variability than 
does the average of many viewpoints reflected in the Homunculus, though 
other factors should not be ignored.

\begin{figure}  
  \includegraphics[angle=0,scale=0.60]{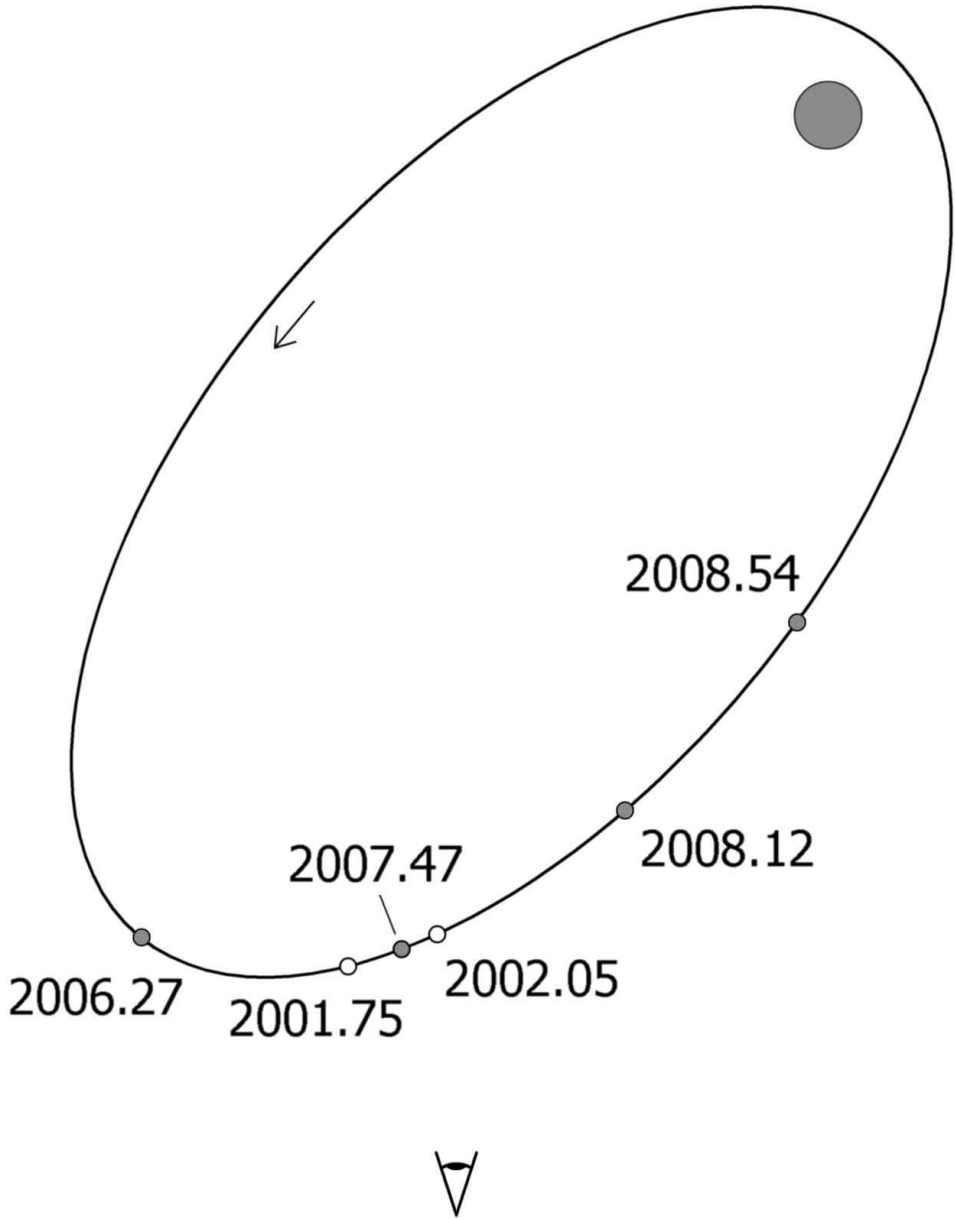}
  \caption{Orbital locations of the  secondary star at times 
     of the spectroscopy discussed here.  The semi-major axis is about 
     17 AU and the eccentricity is assumed to be 0.85 (most authors favor 
     the range 0.75--0.90), with periastron in early July 2003 and late 
     January 2009.  Our viewpoint is below the figure, except that the 
     orbital plane is inclined by roughly $45^{\circ}$.  The apsidal 
     orientation shown here is a compromise between estimates by 
     \citet{kiorbit2001} and \citet{aoorbit2008}, based on X-ray fluxes 
     around apastron;  it is uncertain by at least $\pm  20^{\circ}$. }
  \label{orbitfig}
  \end{figure}

    Since $\eta$ Car's spectroscopic events have attracted attention 
in recent years, there is a natural tendency to ascribe observed 
variability to  the 5.5-year cycle, the wind-wind interaction, and
the influence of the companion star.  But in this case there are good 
arguments against that response.  According to nearly all proposed models 
for the orbit,  in 2006--2007 the wind-wind interaction should have 
been minimal because the companion star was then located far from the 
primary.  During a 3-year interval around apastron, the separation is 
 more than 20 AU (Fig.\ \ref{orbitfig}).  This implies the 
following points for conventional models:       
\begin{enumerate}
  \item Tidal effects are weaker than those at periastron by a factor 
  of order 500.  
  \item If the primary mass loss rate and line-of-sight wind speed are 
  roughly $10^{-3} \ M_{\odot}$ yr$^{-1}$ and wind speeds along the line of sight
  typically 500--600 km s$^{-1}$, then the optical depth for 
  Thomson scattering at 
  $r \gtrsim 20$ AU is small, of the order of 0.1.  
  \item Optical depths for continuum absorption are expected to be even smaller.  
  A detailed model is beyond the scope of this paper, but this assertion 
  is based on two generalities:   Standard processes give  
  $\tau_\mathrm{abs}  <  \tau_\mathrm{sc}$ at the likely densities and 
  temperatures, and  one can also judge the matter from the sizes of 
  photospheres that produce suitable continua.  The visual and near-UV 
  photosphere in the primary wind cannot be very much larger than the circle for the primary 
  in Fig.\  \ref{orbitfig},  $R = 1$ AU;  otherwise its emergent continuum 
  would be too cool.  Therefore the optical depth for continuum 
  absorption must be quite small at $r \sim 20$ AU.  In conventional 
  models the broad-emission-line spectrum originates mostly within 
  $r \lesssim 6$ AU  \citep{Hill01,fos95}, where the secondary star and  
  wind-wind interaction can have at most only a minor indirect influence.
  \item The radiation density is too high for appreciable circumstellar
    dust in this region.  Likely grain formation distances are 150--300 AU
   \citep{dh97}. \citet{Kashi08b} have suggested that dust can form close to 
   the star, but also note that more than 170 days from periastron dust can 
   form only at large distances.  In any case as explained earlier dust, 
   formation cannot explain the photometric decline. 
  \item Along relevant photon paths (see point 6 below), optical depths 
  through the fast, hot, low-density secondary wind are even smaller than 
  those in the primary wind.  
  \item The radiation density is too high for appreciable circumstellar 
  dust in this region.  Likely grain formation distances are 150--300 AU 
  \citep{dh97}.      
  \item When viewing Fig.\ \ref{orbitfig}, it is important to bear in mind 
  the 45{\arcdeg} inclination of the orbit plane.  At 2008.54, for instance, 
  our line of sight to the primary star did not pass close to the secondary 
  star even though the figure might give that impression.  Between 2005.5 
  and 2008.0 our projected view of the system would have changed slowly 
  and continuously -- at least in conventional models.  The same 
  statement applies to expected line-of-sight column densities.    
  \item At visual and near-UV wavelengths, the shocked region between the 
  two winds should be practically transparent for absorption and line 
  formation.  Its high temperatures, $T > 10^6$ K, greatly weaken free-free 
  and other absorption opacities at these wavelengths.  Figures depicting 
  shock simulations such as \citet{pc02}, \citet{aoorbit2008}, and 
  \cite{Park09} do not depict the visual appearance; 
  indeed the shocked regions are expected to emit less than 0.01\% of 
  $\eta$ Car's total red-to-near-UV light.    
  \item  Flow simulations show dramatic spiral patterns in the shocked 
  structure around periastron \citep{aoorbit2008,Park09,2009MNRAS.396.1308G}.
  During most of the orbit, however, the secondary star 
  moves at speeds less than 50 km s$^{-1}$, much slower than either wind;  
  so the flow structure then approximates an axial symmetry around 
  the the star--star axis, with only a modest tilt angle and negligible 
  spirality. 
  \item Near periastron the colliding wind structure tends to become 
  unstable and chaotic \citep{2006ApJ...640..474M,Park09}, 
  but this effect is far less serious at larger distances and lower 
  densities. 
  \item The hot secondary star is thought to ionize the pre-shock zones in the 
  primary wind \citep{davidson97,2008AJ....135.1249H,mehner2010}.
  This probably  affects the \ion{He}{1} P Cyg features described in {\S}3, 
  but it doesn't explain why the column densities and speeds should differ 
  between two cycles.  At most visual and near-UV wavelengths, continuum 
  emission from this photoionized zone should be about two orders of 
  magnitude weaker than that of the primary wind.  In the Balmer continuum 
  it might account for almost 10\% of the total, but this contribution 
  should be quite steady, and insensitive to local gas densities etc.
\end{enumerate} 
In summary, the above points plus the persistence of the spectroscopic changes
supports variations in the primary's wind as opposed to  a wind-wind interaction 
interpretation of the mid-cycle irregularities.   We cannot absolutely 
rule out an explanation of that type, but the obvious alternative 
is far simpler -- i.e., that the observations indicate fluctuations in 
the primary wind.   Similar arguments apply to unconventional models
wherein the secondary star plays the leading role, e.g.\ see  Soker (2007) and 
Kashi and Soker (2009a).

\begin{figure}   
  \includegraphics[angle=270,scale=0.8]{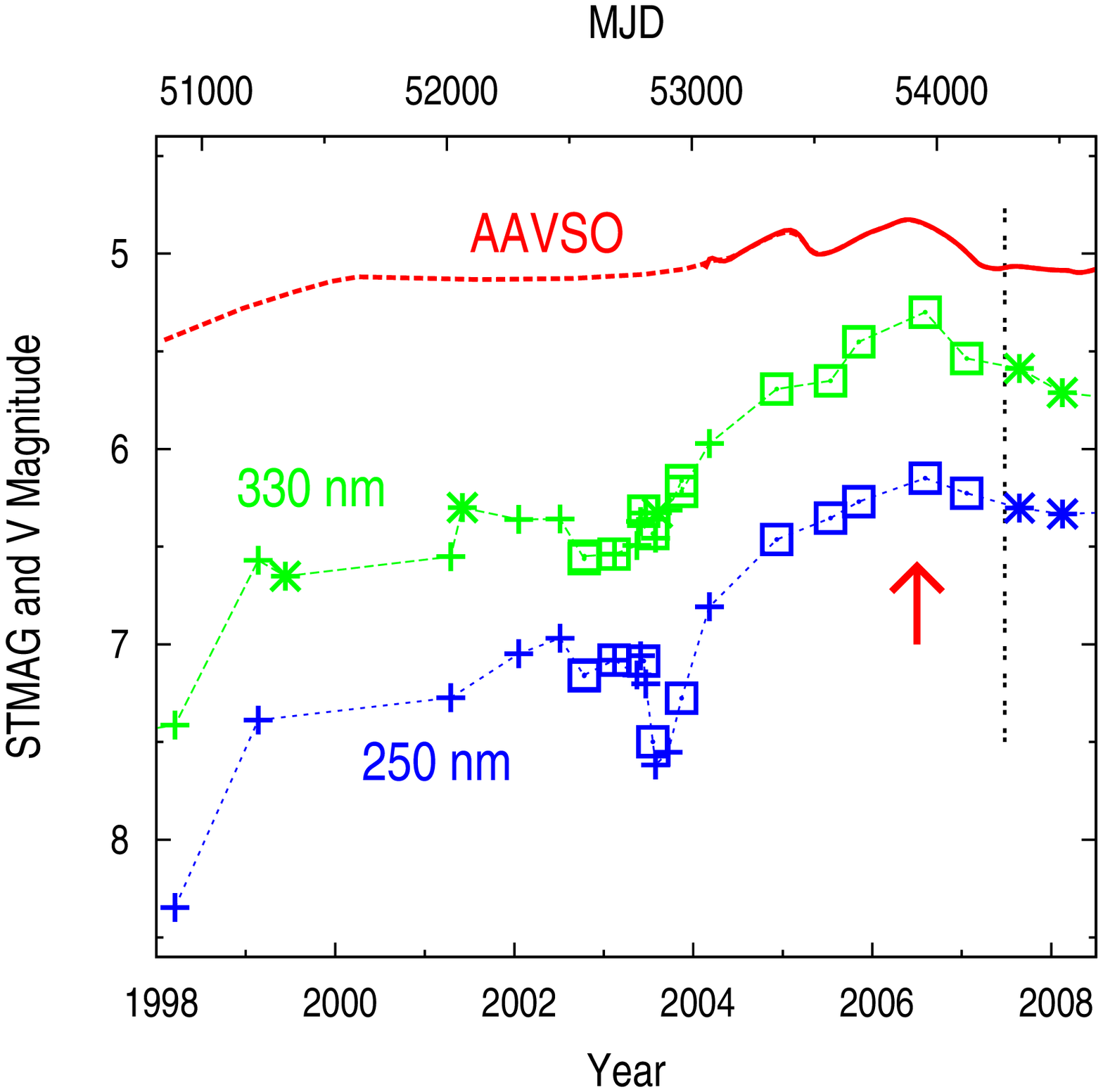}
  \caption{Recent photometric behavior of $\eta$ Car.  The curve marked ``AAVSO'' 
     is a smoothed representation of ground-based V-magnitudes for 
     the Homunculus Nebula plus the star \citep{aavsodata}.  
     Other data shown here include only the central star -- or rather
     its opaque wind --  observed with HST in the UV. 
     Crosses (+) show photometry  synthesized from STIS/CCD 
     spectra near  $\lambda \sim$ 330 and 250 nm 
     \citep{2006AJ....132.2717M}.  Boxes ($\square$) were measured in 
     ACS/HRC images (\citet{2006AJ....132.2717M} and this paper).
     Asterisks ($\ast$) are photometry from WFPC2 images.  These  
     HST points are STMAG values \citep{stmag86}.   An arrow marks 
     the time when ground-based photometry indicated an unusual 
     decrease in brightness.  The vertical dotted line indicates 
     the time of the GMOS spectrum discussed in this paper.}      
  \label{photom}
\end{figure}

The decline in apparent brightness in 2006--2007  was temporary. 
The flux at all wavelengths subsequently leveled off and HST/WFPC2 
photometry after the 2009 event (to be discussed in a forthcoming paper 
on the 2009 event) 
shows that the central star resumed its decades long increase 
in brightness. Furthermore the multi-wavelength visual photometry, and 
the near-UV flux ratio (Fig. \ref{photom})  show that
increased extinction by dust cannot explain the decline.  Allowable 
explanations for the observed decrease in brightness are thus limited. 
One feasible possibility is a decrease in the Balmer continuum emission 
which could be due a change in the stellar wind,
assuming a simple dependence of the continuum on the wind density.

The decrease in the H$\delta$ P Cygni absorption between 2001 and 2007 
suggests a decrease in density of the primary star's wind between the 
two cycles at about the same phase in the orbit. If the density $\rho(r)$ 
of an opaque wind decreases, then its photosphere tends to move inward. 
At constant luminosity (generally assumed for $\eta$ Car), this raises 
the characteristic temperature T$_{c}$ of the emergent radiation. Since 
T$_{c}$ is well above 10000$\arcdeg$ K, in this case, the brightness 
therefore tends to decrease at wavelengths greater than 300nm as the 
radiation shifts further in the UV. This is consistent with the observed 
photometric decrease in 2006--2007. Crude estimates suggest that a density 
decrease on the order of 20\% would suffice. The change in the He I profiles 
also show a corresponding decrease in the wind speed.  The primary's 
wind thus became both slower and less dense mid-cycle. The Balmer
and He I profiles show little change between the GMOS mid 2007 spectrum 
and the spectrum from early 2008. The primary's wind thus stayed in this 
slower, less dense state up until the time of the spectrum from  mid 2008, 
just prior to  the onset of the 2009 event.

A possible explanation for these mid-cycle changes in the stellar wind 
may be related to $\eta$ Car's  aspherical wind, confirmed with near-infrared interferometry \citep{Weigelt07}.  Our line of sight to 
$\eta$ Car is at an angle midway between its fast, dense polar outflow
and its slower, lower density wind in the equatorial region 
\citep{2003ApJ...586..432S}.  A small shift in $\eta$ Car's 
latitude-dependent wind could thus account for the observed spectroscopic 
changes with our line of sight intercepting more of the lower latitude 
wind from mid 2006 to mid 2008.  The He I double P Cygni absorption profile 
in the 2006 UVES spectrum is intriguing in this connection, but with doubts 
about the slit position, its interpretation is uncertain. 

\citet{Kashi09b} have suggested that small changes in the primary's wind can explain
the short term variations observed in the X-ray light curve \citep{Corc05} at
apastron and propose that larger variations may have been responsible for the much
shorter X-ray minimum during the recent 2009 event \citep{Pian09}. However we note that there 
were no large fluctuations in the X-ray flux during the time of the photometric
decline and our observed spectral changes from $\approx$ late 2006 to at least early  
2008, although the observed 20\% change in the wind velocity may result in only a
7\% decrease in the X-ray flux \citep{Kashi09b}.

We also noted mid-cycle changes in 2000-2002 although they were somewhat 
different and did not occur at  the same phase in the orbit. These 
small mid-cycle fluctuations may be part of a longer term  secular trend 
perhaps related to the star's rapid brightening in the past decade as it 
recovers from the giant eruption.  Most recent work on $\eta$ Car has 
focused on its spectroscopic events, but mid-cycle observations  merit 
more attention. Mid-cycle variability can provide important information on the primary
star's intrinsic instability that the spectroscopic events do not.

We thank Jean-Rene Roy, Deputy Director and Head of Science Operations 
of the Gemini Observatory, for granting us director's discretionary time on 
short notice to obtain spectra of $\eta$ Car in 2007.  We also
thank the staff and observers of the Gemini South observatory in La Serena for
their help and support in putting together a successful observing program.
We are also acknowledge use of the AAVSO International Database.


	\begin{deluxetable}{lcccc}       
	\tablewidth{0pt}
	\tablecaption{Measured spatial sampling efficiency of 0.3{\arcsec} virtual 
	  aperture\tablenotemark{a} \ for a point source \label{caltab}}
	\tablecolumns{3}
	\tablehead{ \colhead{HST camera} 
	      & \colhead{F250W, F255W}  &  \colhead{F330W, F336W} }
	\startdata
	ACS (HRC)\tablenotemark{b}  & 0.594 $\pm$ 0.013 & 0.625 $\pm$ 0.010 \\
	WFPC2 (PC)\tablenotemark{c} & 0.567 $\pm$ 0.030 & 0.669 $\pm$ 0.048 \\  
	\enddata
	\tablenotetext{a}{ Weighting function  \ $1 - r^2/R^2$ \ with 
	   \ $R = 0.15{\arcsec}$.  }     
	\tablenotetext{b}{ \citet{2006AJ....132.2717M} }.
	\tablenotetext{c}{ The r.m.s.\ errors quoted here are relevant to 
	   comparisons between ACS and WFPC2 data, but maybe pessimistic 
	   regarding brightness {\it variations} with a given filter.}

	\end{deluxetable}


	\begin{deluxetable}{lllrrrr}   
	\tablewidth{0pt}
	\tabletypesize{\tiny}
	\tablecaption{Photometry Results\label{phottab}}
	\tablecolumns{7}
	\tablehead{
	&&&\colhead{Exp Time}&\colhead{Flux}\\
	\colhead{Dataset}&\colhead{MJD}&\colhead{Year}&\colhead{(sec)}&\colhead{Density\tablenotemark{c}}&
	\colhead{Magnitude\tablenotemark{a}}&\colhead{Average\tablenotemark{b}}
	}
	\startdata
	\multicolumn{7}{c}{ACS HRC/F250W Filter}\\
	\tableline\\
	j9p603onq&54120.426&2007.055&1.2&1.189&6.212&6.226$\pm$0.010\\
	j9p603opq&54120.426&2007.055&1.2&1.159&6.239&\nodata\\
	j9p603orq&54120.430&2007.055&1.2&1.174&6.225&\nodata\\
	j9p603otq&54120.430&2007.055&1.2&1.172&6.228&\nodata\\
	\tableline\\
	\multicolumn{7}{c}{WFPC2 HRC/F255W Filter}\\
	\tableline\\
	ua140109m&54335.188&2007.643&0.6&1.094&6.302&6.302\\
	ua140209m&54510.797&2008.124&0.6&1.064&6.333&6.333\\
	\tableline\\
	\multicolumn{7}{c}{ACS HRC/F330W Filter}\\
	\tableline\\
	j9p603obq&54120.410&2007.055&0.4&1.336&6.085&6.087$\pm$0.003\\
	j9p603odq&54120.410&2007.055&0.4&1.338&6.084&\nodata\\
	j9p603ofq&54120.414&2007.055&0.4&1.329&6.091&\nodata\\
	j9p603ohq&54120.414&2007.055&0.4&1.334&6.087&\nodata\\
	\tableline\\
	\multicolumn{7}{c}{WFPC2 HRC/F336W Filter}\\
	\tableline\\
	ua140101m&54335.121&2007.643&0.1&1.273&6.138&6.138$\pm$0.020\\
	ua140102m&54335.125&2007.643&0.1&1.312&6.105&\nodata\\
	ua140103m&54335.121&2007.643&0.1&1.253&6.155&\nodata\\
	ua140104m&54335.125&2007.643&0.1&1.254&6.154&\nodata\\
	ua140201m&54510.770&2008.124&0.1&1.154&6.245&6.262$\pm$0.019\\
	ua140202m&54510.773&2008.124&0.1&1.135&6.263&\nodata\\
	ua140203m&54510.773&2008.124&0.1&1.151&6.247&\nodata\\
	ua140204m&54510.777&2008.124&0.1&1.105&6.292&\nodata\\
	\enddata
	\tablenotetext{a}{Magnitude on the STMAG system.}
	\tablenotetext{b}{The average and sigma of individual measurements in
	  a set of exposures taken within a day of each other.}
	\tablenotetext{c}{STMAG flux units are 10$^{-11}$ erg/cm$^2$/s/\mbox{\AA}.}
	\end{deluxetable}

	\begin{deluxetable}{lrrr}
	\tablewidth{0pt}
	\tabletypesize{\tiny}
	\tablecaption{Spectral Observations\label{spectab}}
	\tablecolumns{4}
	\tablehead{
	\colhead{Root}&&&\colhead{Exp Length}\\
	\colhead{Name}&\colhead{Year}&\colhead{Grating}&\colhead{(sec)}\\
	}
	\startdata
	\multicolumn{4}{c}{Gemini South GMOS}\\
	\tableline\\
	S20070616S0045&2007.45&B1200&5.5\\
	S20070616S0046&2007.45&B1200&5.5\\
	S20070618S0011&2007.45&B1200&5.5\\
	S20070618S0012&2007.45&B1200&5.5\\
	S20070630S0026&2007.49&R831&3.5\\
	S20070630S0035&2007.49&R831&1.5\\
	S20070630S0040&2007.49&R831&4.5\\
        S20080211S0087&2008.11&B1200&5.5\\
        S20080211S0088&2008.11&B1200&5.5\\
        S20080718S0068&2008.54&B1200&5.5\\
        S20080718S0069&2008.54&B1200&5.5\\
	\tableline\\
	\multicolumn{4}{c}{HST STIS}\\
	\tableline\\
	o6ex03010&2001.75&G750M&  6.0\\
	o6ex03020&2001.75&G750M&  0.6\\
	o6ex03060&2001.75&G430M& 36.0\\
	o6ex030e0&2001.75&G430M& 36.0\\
	o6ex030f0&2001.75&G430M&  3.0\\
	o6ex02010&2002.05&G750M&  6.0\\
	o6ex02020&2002.05&G750M&  0.4\\
	o6ex02030&2002.05&G430M& 18.0\\
	o6ex020u0&2002.05&G430M&  8.0\\
	o6ex020v0&2002.05&G430M& 36.0\\
	\tableline\\
	\multicolumn{4}{c}{VLT UVES}\\
	\tableline\\
        UVES.2006-04-09&2006.27&CD\#2&60.0\\
        T02:41:56.828\\
	\enddata
	\end{deluxetable}

\begin{deluxetable}{lrrr} 
    \tablewidth{0pt}
    \tabletypesize{\tiny}
    \tablecaption{Instrument Parameters\label{instab}}
    \tablecolumns{4}
     \tablehead{ 
     &\colhead{Slit Width}&\colhead{Slit Angle\tablenotemark{a}}
     &\colhead{Spectral}\\ &\colhead{(arcseconds)}&\colhead{(deg)}& 
     \colhead{R = $\frac{lambda}{\Delta\lambda}$}
     }
											              \startdata 
   Gemini South&0.50&+160&3744\tablenotemark{b}\\
   GMOS&&&4396\tablenotemark{c}\\
   \tableline\\
   HST STIS&0.10&+165\tablenotemark{d}&15350\tablenotemark{b}\\
   &&-82\tablenotemark{d}&13400\tablenotemark{c}\\
   \tableline\\
   VLT UVES&0.40&+160&80000\tablenotemark{b}\\
   \enddata
   \tablenotetext{a}{The slit angle is measured from north through east.
   All slits are targeted on the central star.}
   \tablenotetext{b}{R for spectra blueward of 500 nm.}
   \tablenotetext{c}{R for spectra redward of 500 nm.}
   \tablenotetext{d}{Slit orientation was 
  +165 degrees in October 2001 and -82 degrees in January 2002.}
  \end{deluxetable}


\begin{thebibliography}{}  
	\bibitem[Altamore et al.(1986)]{altamore86} Altamore, A., Baratta, G.~B., 
	  Cassatella, A., Rossi, L., \& Viotti, R.\ 1986, New Insights in 
	  Astrophysics: Eight Years of UV Astronomy with IUE, 303 
	\bibitem[Biretta et al.(1996)]{WFPC2_Handbook} Biretta, J. A., et al. 1996, 
	  WFPC2 Instrument Handbook, Version 4.0 (Baltimore: STScI)
	\bibitem[Cassatella, Giangrande, \& Viotti(1979)]{cassatella79} Cassatella, A.,   Giangrande, A., \& Viotti, R.\ 1979, \aap, 71, L9
	\bibitem[Corcoran (2005)]{Corc05} Corcoran, M. F., 2005, \aj, 129, 2018
        \bibitem[Damineli et al.(1997)]{ad97} Damineli, A., Conti, P.S., 
	  \& Lopes, D.F. 1997, New Astronomy, 2, 107  
        \bibitem[Damineli et al.(2008)]{ad2008} Damineli, A., Hillier, D.~J., 
	   Corcoran, M.~F., et al. 2008, \mnras, 386, 2330  
	\bibitem[Davidson et al.(1995)]{fos95} Davidson, K., Ebbets, D., Weigelt, G., 
	  Humphreys, R.M., Hajian, A.R., Walborn, N.R., \& Rosa, M.\ 1995, 
	  \aj, 109, 1784 
        \bibitem[Davidson(1997)]{davidson97} Davidson, K. 1997, 
           New Astronomy, 2, 387 
        \bibitem[Davidson \& Humphreys(1997)]{dh97} Davidson, K., \& 
	     Humphreys, R.M. 1997, \araa, 35, 1  
	\bibitem[Davidson et al.(1999a)]{1999AJ....118.1777D} Davidson, K., 
	  et al.\ 1999a, \aj, 118, 1777

	\bibitem[Davidson et al.(1999b)]{stis99} Davidson, K., Ishibashi, K., 
	  Gull, T.R., \& Humphreys, R.M. 1999b, Eta Carinae at the Millenium, 
	  ASP Conf.\ Ser.\ 179 (ed.\ J.A.\ Morse, R.M.\ Humphreys, \& 
	  A.\ Damineli), 227   
        \bibitem[Davidston et al.(2001)]{homunc01} Davidson, K., Smith, N., 
	  Gull, T.~R., Ishibashi, K., \& Hillier, D.~J. 2001,  
          \aj, 121, 1569   
	\bibitem[Davidson(2004)]{Davidson04} Davidson, K.  2004, 
	  STScI Newsletter, Spring 2004, 1 
	\bibitem[Davidson et al.(2005)]{2005AJ....129..900D} Davidson, K., 
	  et al.\  2005, \aj, 129, 900
	\bibitem[DiScala \& Jones (2008)] {aavsodata} DiScala, G., \& 
	  Jones, R.~W.\ 2008, Observations from the AAVSO International 
	  database, private communication.
	\bibitem[Feinstein \& Marraco(1974)]{1974A&A....30..271F} Feinstein, A., 
	  \& Marraco, H.~G.\ 1974, \aap, 30, 271 
	\bibitem[Feinstein(1967)]{1967Obs....87..287F} Feinstein, A.\ 1967, 
	  The Observatory, 87, 287 
	\bibitem[Fernandez-Lajus et al.(2003)]{2003IBVS.5477....1F} Fernandez-Lajus, 
	  E., Gamen, R., Schwartz, M., Salerno, N., Llinares, C., Farina, C., 
	  Amor{\'{\i}}n, R., \& Niemela, V.\ 2003, Information Bulletin on 
	  Variable Stars, 5477, 1
	\bibitem[Fernandez-Lajus et al.(2009)]{Lajus09}Fernández-Lajús, E., Fariña, C., 
	Torres, A. F., Schwartz, M. A., Salerno, N., Calderón, J. P., von Essen, C., 
	Calcaferro, L. M., Giudici, F., Llinares, C., \& Niemela, V., \ 2009, \aap, 493, 1093
        \bibitem[Gull et al.(2009)]{2009MNRAS.396.1308G} Gull, T.~R., 
          et al.\ 2009, \mnras, 396, 1308
        \bibitem[Hillier \& Allen(1992)]{ha92} Hillier, D.J., \& Allen, D.A. 
           1992, \aap, 262, 153 
	\bibitem[Hillier et al.(2001)]{Hill01} Hillier, D.J., Davidson, K., Ishibashi, K.,
	\& Gull, T.~R. \ 2001, \apj, 553, 837
        \bibitem[Hillier et al.(2006)]{Hill06} Hillier, D.J., Gull, T., 
	   Nielsen, K., et al. 2006, \apj,, 642, 1098  
	\bibitem[Hofmann \& Weigelt(1988)]{gw88} Hofmann, K.-H., \& Weigelt, G. 
	  1988, \aap, 203, L21
	\bibitem[Holtzman et al.(1995)]{1995PASP..107..156H} Holtzman, J.~A., et al.\ 
	  1995, \pasp, 107, 156 
	\bibitem[Humphreys \& Stanek(2005)]{ASP332} Humphreys, R.~M., \& Stanek, 
	  K.\ (eds.) 2005, ASP Conf.\ Ser.\ 332, The Fate of the Most Massive Stars
	\bibitem[Humphreys et al.(2008)]{2008AJ....135.1249H} Humphreys, R.~M., 
	  Davidson, K., \& Koppelman, M.\ 2008, \aj, 135, 1249
	\bibitem[Ishibashi(2001)]{kiorbit2001} Ishibashi, K., Eta Carinae and 
	  Other Mysterious Stars, ASP Conf.\ Ser.\ 242 (ed.\ T.R.\ Gull,   
	  S.\ Johansson, \& K.\ Davidson), 53
	\bibitem[Kashi \& Soker (2008a)]{Kashi08a}Kashi, A. \& Soker, N.\ 2008a \mnras, 390, 1751
        \bibitem[Kashi \& Soker (2008b)]{Kashi08b}Kashi, A. \& Soker, N.\ 2008b New Astronomy, 13, 569
	\bibitem[Kashi \& Soker (2009a)]{Kashi09a}Kashi, A. \& Soker, N.\ 2009a New Astronomy,        14, 11  
	\bibitem[Kashi \& Soker (2009b)]{Kashi09b}Kashi, A. \& Soker, N.\ 2009b \apj, 701, L59
        \bibitem[Koornneef et al.(1986)]{stmag86} Koornneef, J., Bohlin, R., 
          Buser, R., Horne, K., \& Turhshek, D. 1986, Highlights Astron., 
          7, 833 
	\bibitem[Martin(2005)]{2005ASPC..332..111M} Martin, J.~C.\ 2005, The Fate  
	  of the Most Massive Stars, ASP Conf.\ Ser.\ 332, 111 
	\bibitem[Martin \& Koppelman(2004)]{2004AJ....127.2352M} Martin, J.~C., \& 
	  Koppelman, M.~D.\ 2004, \aj, 127, 2352
	\bibitem[Martin et al.(2006a)]{2006ApJ...640..474M} Martin, J.~C., Davidson, 
	  K., Humphreys, R.~M., Hillier, D.~J., \& Ishibashi, K.\ 2006a, \apj, 
	  640, 474
	\bibitem[Martin et al.(2006b)]{2006AJ....132.2717M} Martin, J.~C., Davidson, 
	  K., \& Koppelman, M.~D.\ 2006b, \aj, 132, 2717
	\bibitem[Martin et al.(2006c)]{2006PASP..118..697M} Martin, J.~C., Davidson, 
	  K., Hamann, F., Stahl, O., \& Weis, K.\ 2006c, \pasp, 118, 697 
	\bibitem[Mattei \& Foster(1998)]{1998IAPPP..72...53M} Mattei, A., \& 
	  Foster, G.\ 1998, International Amateur-Professional Photoelectric 
	  Photometry Communications, 72, 53 
        \bibitem[Mehner et al.(2010)]{mehner2010} Mehner, A., Davidson, K., 
	  Ferland, G.~J., \& Humphreys, R.~M. 2010, \apj, in press 
	 \bibitem[Nielsen et al (2009)]{Nielsen09}Nielsen, K. E., Kober, G. Vieira, 
	 Weis, K., Gull, T. R., Stahl, O., \& Bomans, D. J.\ 2009 \apjs, 181, 473 
	\bibitem[O'Connell(1956)]{1956VA......2.1165O} O'Connell, D.~J.~K.\ 1956, 
	  Vistas in Astronomy, 2, 1165 
	\bibitem[Okazaki et al.(2008)]{aoorbit2008} Okazaki, A.T., Owocki, S.P., 
	  Russell, C.M.P., \& Corcoran, M.F.\ 2008, \mnras, 388, L39  
	\bibitem[Osterbrock \& Ferland(2006)]{ofbook05} Osterbrock, D.E., 
	  \& Ferland, G.J.\ 2006, {\it Astrophysics of Gaseous Nebulae and 
	  Active Galactic Nebulae} 
	\bibitem[Parkin et al.(2009)]{Park09}Parkin, E. R., Pittard, J. M., Corcoran, M.
	F., Hamaguchi, K. \& Stevens, I. R. \ 2009, \mnras, 394, 1758 
	\bibitem[Pian et al. (2009)]{Pian09}Pian, E., Campana, S., 
	  Chincarini, G., Corcoran, M. F., Hamaguchi, K., Gull, T., 
	  Mazzali, P. A., Thoene, C. C., Morris, D., \& Gehrels, N.\ 2009, 
	  arXiv:0908.2819   
        \bibitem[Pittard \& Corcoran(2002)]{pc02} Pittard, J.M., \& Corcoran, M.F. 
           2002, \aap, 383, 636   
	\bibitem[Sirianni et al.(2005)]{2005PASP..117.1049S} Sirianni, M., et al.\ 
	  2005, \pasp, 117, 1049 
	\bibitem[Smith et al.(2003)]{2003ApJ...586..432S} Smith, N., Davidson, K.,
	  Gull, T.~R., Ishibashi, K., \& Hillier, D.~J.\ 2003, \apj, 586, 432 
	\bibitem[Smith et al.(2004)]{Smith04}Smith, N. et al,\ 2004, \apj, 605, 405 
	\bibitem[Soker(2007)]{Soker07}Soker, N. 2007, \apj, 661, 490 
	\bibitem[Sterken et al.(1999)]{1999A&A...346L..33S} Sterken, C., Freyhammer, 
	  L., Arentoft, T., \& van Genderen, A.~M.\ 1999, \aap, 346, L33
	\bibitem[van Genderen et al. (1999)]{vanG99}van Genderen, A. M., Sterken, C., de Groot,        M., \& Burki, G. \ 1999, \aap, 343, 847
	\bibitem[van Genderen \& Sterken (2004)]{vanG04}van Genderen, A. M. \& Sterken, C,\
	2004, \aap, 423, L1   
	\bibitem[Viotti et al.(1989)]{viotti89} Viotti, R., Rossi, L., Cassatella, A., 
	  Altamore, A., \& Baratta, G.~B.\ 1989, \apjs, 71, 983  
	\bibitem[Weigelt \& Ebersberger(1986)]{gw86} Weigelt, G., \& 
	  Ebersberger, J. 1986, \aap, 163, L5  
	\bibitem[Weigelt et al (2007)]{Weigelt07}Weigelt, G., Kraus, T. Driebe, T. et al \
	2007, \aap, 464, 87
	\bibitem[Whitelock et al.(2004)]{2004MNRAS.352..447W} Whitelock, P.~A., 
	  Feast, M.~W., Marang, F., \& Breedt, E.\ 2004, \mnras, 352, 447 
	\bibitem[Zanella et al.(1984)]{1984A&A...137...79Z} Zanella, R., Wolf, B., 
	  \& Stahl, O.\ 1984, \aap, 137, 79
	\end{thebibliography}
\end{document}